# Controlled complete suppression of single-atom inelastic spin and orbital cotunnelling

B. Bryant[1], R. Toskovic[1], A. Ferrón[2,3], J. L. Lado[2], A. Spinelli[1], J. Fernández-Rossier[2], and A. F. Otte[1*]

[1]Department of Quantum Nanoscience, Kavli Institute of Nanoscience, Delft University of Technology, Lorentzweg 1, 2628 CJ Delft, The Netherlands

[2]International Iberian Nanotechnology Laboratory (INL), Avenida Mestre José Veiga, 4715-310 Braga, Portugal

[3]Instituto de Modelado e Innovación Tecnológica (CONICET-UNNE), Avenida Libertad 5400, W3404AAS, Corrientes, Argentina

* a.f.otte@tudelft.nl

**The inelastic portion of the tunnel current through an individual magnetic atom grants unique access to read out and change the atom's spin state[1], but it also provides a path for spontaneous relaxation and decoherence[2,3]. Controlled closure of the inelastic channel would allow for the latter to be switched off at will, paving the way to coherent spin manipulation in single atoms. Here we demonstrate complete closure of the inelastic channels for both spin and orbital transitions due to a controlled geometric modification of the atom's environment, using scanning tunnelling microscopy (STM). The observed suppression of the excitation signal, which occurs for Co atoms assembled into chain on a $Cu_2N$ substrate, indicates a structural transition affecting the $d_z^2$ orbital, effectively cutting off the STM tip from the spin-flip cotunnelling path.**

Cotunnelling is a two-step process that may happen whenever an electron-confining island (e.g. an atom, molecule or quantum dot) is weakly coupled to two electrodes: an electron hops from the first electrode to the island and another electron hops from the island to the second electrode. If the final state has a different energy than the initial state, the cotunnelling process is inelastic and can only occur if sufficient bias voltage is applied. This effect has proven highly useful in probing e.g. molecular vibrations[4,5], excitations of quantum dots[6] and carbon nanotubes[7], and spin-flip excitations on molecules[8–10] and single atoms[1,11–13]. Inelastic processes involving a spin-flip can only occur when tunnelling into a singly-occupied level[14].

Atomic assembly of magnetic nanostructures allows for fine control of parameters such as single-atom magneto-crystalline anisotropy[15,16] and spin coupling[17,18]. Here we show, using both spin and orbital inelastic excitation spectra, that strain induced by neighbouring atoms can lead to a change in the orbital filling, resulting in complete closure of an inelastic cotunnelling channel. For Co atoms assembled into chains we observe a complete suppression of spin excitations, accompanied by a partial suppression of newly observed orbital excitations at higher energy.

Fig. 1 shows an STM topographic image and inelastic tunnelling spectroscopy (IETS) measurements of a dimer of Co atoms, constructed using atom manipulation on a $Cu_2N$ surface, and also of a Co trimer and tetramer constructed as extensions of the dimer. All nanostructures were constructed with two unit-cell spacing along the direction of the neighbouring N atoms (x-axis). The dimer spectra differ from the single Co case[19], but can be fitted with simulated spin excitation spectra[20] by including an antiferromagnetic exchange coupling of 2 meV (supplementary Fig. S1). A remarkable effect occurs for chains of 3 atoms or longer (Fig. 1 b and c): the end atoms of the chains show spin excitation steps around 8 and 13 mV, but no steps are seen on the inner atoms, even up to 75 meV (see supplementary Fig. S2). The theoretical maximum energy for a single $\Delta S=1$ spin flip transition is $\Delta_{max}= \lambda \Delta S L_{max}$ = 66 meV[21], between orbital states split by the spin-orbit coupling $\lambda$ = 22 meV with maximum unquenched orbital momentum $L_{max}$ =3.

Fig. 2b shows IETS spectra of a Co hexamer on $Cu_2N$, again without steps on the inner atoms. Significant changes are also seen in spectroscopy performed at higher bias voltage. In spectra taken on the end atoms up to ± 1 V, a prominent peak at +500 mV is observed (Fig. 2c); single Co atoms show similar spectra (red curve). But for the inner atoms, all spectroscopic features at positive voltage, including the peak at +500 mV, disappear. Also, spectral features at negative bias are more pronounced on the inner atoms. The correspondence between suppression of steps at low voltage and high positive voltage is found for all Co atoms on $Cu_2N$.

The spectral feature at +500 mV may be associated with an inelastic excitation that involves a change in orbital filling, as observed previously in quantum dots[6,22]. A schematic of spin and orbital excitation processes is shown in Fig. 3. Spin IETS transitions involve a single spin flip and thus cost the Zeeman and/or magnetic anisotropy energy of a few meV: electrons tunnel onto and off the same orbital. Orbital IETS transitions involve tunnelling on and off different orbitals, creating an orbital excited state. The energy cost of these transitions will be the crystal field splitting, which is of order several hundred meV[16].

The combined suppression of both spin IETS and orbital IETS transitions can be understood in terms of a simple qualitative model. The $d_z^2$ orbital of the magnetic atom dominates transport between the tip and the atom: the other $d$ orbitals couple only to the substrate. Therefore, all spin and orbital cotunnelling excitations must involve the $d_z^2$ orbital[23]. In case the $d_z^2$ orbital is half-filled (Fig. 3a), spin and orbital excitations are allowed for both voltage polarities. This is the case for single Co atoms and for outer atoms of chains. If the $d_z^2$ orbital is either fully occupied or empty, however, spin excitations are blocked completely and orbital excitations can occur only at negative sample voltage (Fig. 3b), as observed for inner atoms.

The change in filling of the $d_z^2$ orbital may be due to changes in the crystal environment of the Co atom. DFT calculations (Figs. 3c and d) indicate that for inner atoms of Co chains, the N-Co-N angle $\theta$ is almost collinear (175°), whereas for a single Co atom $\theta$ = 150°. This drastic change in structure influences the atom's magneto-crystalline anisotropy. This can be verified by spin IETS spectra, which show that Co atoms assembled into chains show a substantial anisotropy enhancement compared to single atoms. Figs. 4a and b show IETS spectra of a tetramer: as before, on the inner atoms spin excitations are suppressed, as well as positive-bias orbital excitations. Without spin IETS steps on the inner atoms, we cannot directly measure their magnetic anisotropy. However, in a minority of cases (4 out of 20 structures) we observed Co N-row chains in which spin excitations on the inner atoms were not suppressed (Figs. 4c, and supplementary Fig. S3). Significantly, in these cases all atoms also show the prominent orbital IETS peak close to +500 mV (Fig. 4d), reinforcing the correlation of suppression of spin and positive-bias orbital transitions. We believe that in these minority cases all the atoms of the chain have a partially occupied $d_z^2$ orbital.

We can fit simulated spin IETS spectra to all the atoms of the "unsuppressed" chain (Fig. 4c) and the end atoms of the "suppressed" chain (Fig. 4a), by varying only the component $\Lambda_{xx}$ of the anisotropy tensor $\Lambda$[16,24], corresponding to the chain axis ($x$) direction. The anisotropy of the inner atoms of the

"suppressed" chain may be estimated by its effect on the outer atom spectra. We find that all atoms have a larger $\Lambda_{xx}$ than a single Co atom, and that the inner atoms of both chains have a larger $\Lambda_{xx}$ than the outer atoms. Moreover, all atoms of the "suppressed" chain have a larger anisotropy than those of the "unsuppressed" chain. This effect may be understood in terms of changes in the crystal field: as θ approaches 180° the crystal field approaches the high-symmetry linear case, and the magnetic anisotropy increases rapidly[16]. Since the inner atoms of the "suppressed" chain have the largest $\Lambda_{xx}$ we can infer that they have θ closest to 180°. End atoms of chains have smaller θ, and hence smaller $\Lambda_{xx}$, than the inner atoms.

We compared DFT simulations of the magnetization density for different Co structures on $Cu_2N$ (Figs. 3e and f). For a single Co atom the magnetization profile in the xz plane has a $d_{xz} + d_z^2$ character, but for an inner atom of a chain, it shows the perfect fourfold symmetry of the $d_{xz}$ orbital, indicating that the $d_z^2$ orbital is no longer contributing to magnetism. This is due to strong hybridization of the $d_z^2$ orbital with the N orbitals, which occurs since the N-Co-N angle θ is almost 180°, leading to a situation which is equivalent to the picture shown in Fig. 3, in which the Co $d_z^2$ orbital is fully filled. Thus, we have a mechanism whereby a transition to a higher-symmetry crystal environment leads to a modification of the orbital filling, and an effectively fully-filled, non-magnetic $d_z^2$ orbital. This effect is not observed for chains of Fe on $Cu_2N$[2]: these show a smaller θ (supplementary Fig. S4) and hence are not expected to show strong hybridization effects.

We can account for the two different types of N-row structures shown in Fig. 4 by variation in local strain. Single Co atoms on $Cu_2N$ show just a ± 5% variation in anisotropy, similar to Fe atoms and dimers on the same surface, but Co N-row dimers[20] and chains are highly sensitive to strain. Since θ is already close to 180°, a small change in strain caused by subsurface defects can drive a large change in anisotropy, and can also reach the critical value of θ that triggers the transition to a hybridised $d_z^2$ orbital.

In summary, we have demonstrated how modifications in the crystal environment of a single Co atom can result in complete closure of both spin and orbital cotunneling paths. A small variation in an external parameter (strain) can effectively turn spin excitations on and off, similar to the role of a gate voltage changing the occupancy of a quantum dot from odd to even[14]. In the present experiment we can controllably modify this strain by adding an extra atom to a nanostructure (Figs. 1, S3). We foresee that if magnetic nanostructures were assembled on a piezoelectric substrate, it would be possible to control the global strain, providing a method to tune orbital occupancy, and thereby inelastic channel closure, through a separate voltage signal.

**Methods**

Measurements were carried out at 330 mK in ultra-high vacuum (< $2×10^{-10}$ mbar), in a commercial STM system (Unisoku USM-1300S). The $Cu_2N$ substrate was prepared in situ by $N_2$ sputtering of a Cu(100) crystal. Co atoms were evaporated onto the precooled $Cu_2N$ surface. STM tips were prepared by indenting commercial Pt-Ir STM tips into the Cu surface. Co nanostructures were assembled on the $Cu_2N$ surface using vertical atom manipulation. For IETS measurements, differential conductance (d$I$/d$V$) spectra were recorded with a lock-in amplifier, using an oscillation amplitude of 50 μV rms, at a typical conductance values of 1 μS for spectra up to 20 mV and 2 nS for spectra up to 1 V.

Simulated IETS spectra are produced by diagonalization of a spin Hamiltonian that includes nearest-neighbour Heisenberg exchange interaction as well as magnetic anisotropy in terms of a second-order perturbative treatment of the spin-orbit coupling[16,20]. Resulting lineshapes are generated by taking interactions with the substrate electrons into account up to third order[25].

DFT calculation were carried out using Quantum Espresso (QE)[26] and Elk[27]. Structural relaxations were performed using QE with PAW pseudopotentials and PBE exchange correlation functional, for the cells with 3x3 and 3x2 with 4 layers of Cu(100) and 1 layer of N, corresponding to the monomer and the chain. These calculations show a decrease in the N-Co-N angle θ from 175° for the Co chain to 150° for the monomer.

In the case of the monomer, relaxed structures overestimate the Co-Cu coupling[28], thereby reducing the spin. To demonstrate the correlation between θ and the magnetization profile, we therefore show in Fig. 3d the profile for a monomer with reduced N-Co-N angle θ = 110°, for which S=3/2, in agreement with experiment. The magnetization profiles were calculated using all-electron LAPW Elk code, with LDA+U in the fully localized limit and Yukawa scheme[29] with a screening length of 1.88 atomic units.

**Acknowledgements**
This work was supported by the Dutch funding organizations FOM and NWO (VIDI) and by the Kavli Foundation. JFR acknowledges financial support by MEC-Spain (FIS2013-47328-C2-2-P) and Generalitat Valenciana (ACOMP/2010/070 and Prometeo). This work has been financially supported in part by FEDER funds. JLL and JFR acknowledge financial support by Marie-Curie-ITN 607904-SPINOGRAPH. AF acknowledges Marie Curie COFUND (grant number 600375) and CONICET. The authors thank Markus Ternes for providing the software to simulate IETS spectra.

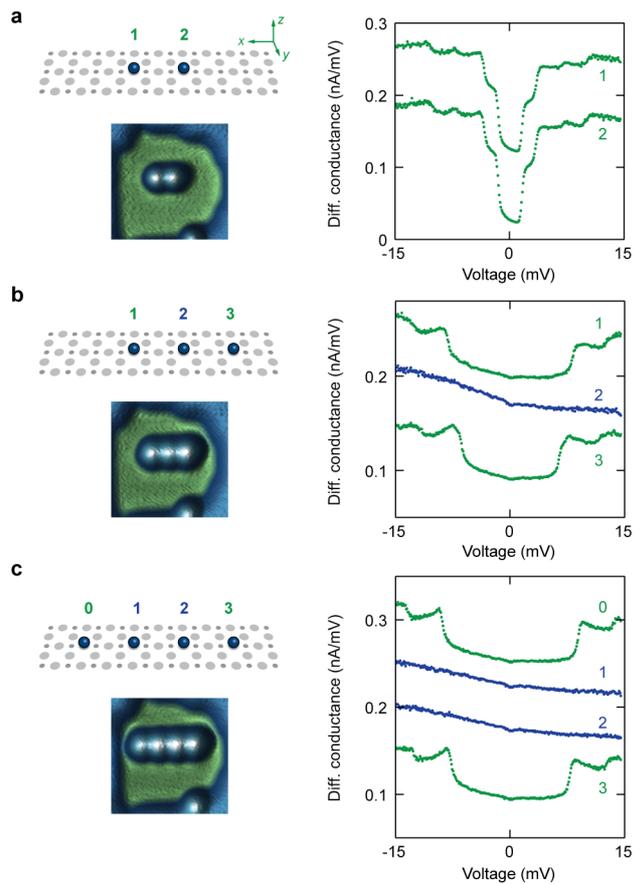

**Figure 1: Suppression of spin inelastic excitations in Co nanostructures.** Structural model, STM topograph and IETS spectra of a Co dimer built by atom manipulation on $Cu_2N$ **(a)**, and a Co trimer **(b)** and tetramer **(c)** built as extensions of the dimer. The spectra of the dimer, and the edge atoms of the trimer and tetramer (shown in green) show spin IETS steps. Spectra of the inner atoms of the trimer and tetramer (blue) show no spin excitations.

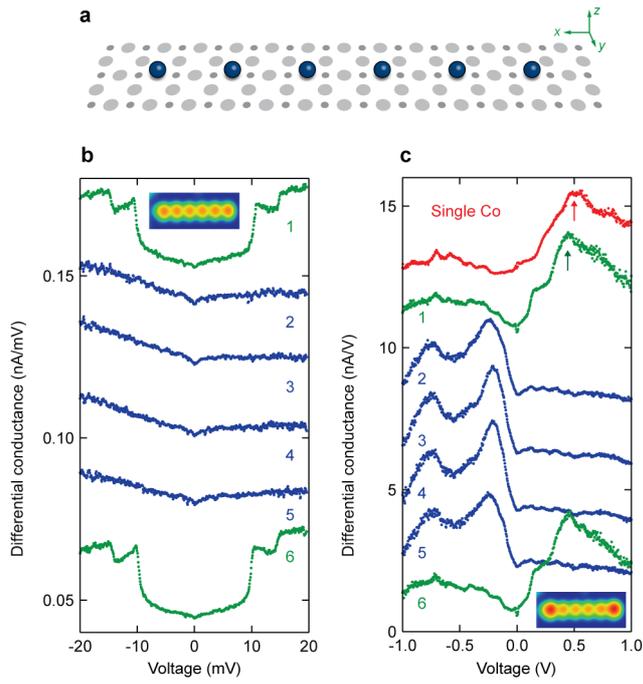

**Figure 2: Orbital and spin excitations in a Co nanostructure (a)** structure of a Co N-row hexamer on Cu₂N **(b)** spin IETS spectra of a Co hexamer, showing suppression of IETS steps on inner atoms. A topographic image at +20 mV is shown (inset) **(c)** higher bias spectra on the same hexamer. The peak at +500 mV sample bias seen on the single Co (indicated) is attributed to an orbital excitation: the same peak is seen on the end atoms of the hexamer, though slightly shifted. In the inner atoms of the hexamer all positive bias spectral features are suppressed. A topographic image at +500 mV is shown (inset).

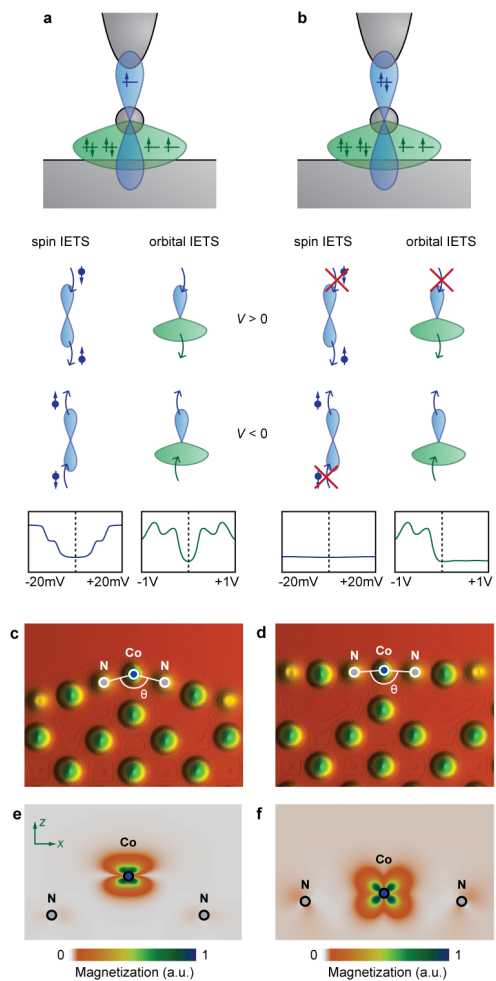

**Figure 3: Suppression of orbital and spin IETS transitions. (a)** Schematic of spin and orbital inelastic transitions. The $d_z^2$ orbital (blue) couples to the tip and the substrate: the other d orbitals (green) couple only to the substrate. Inelastic spin excitations involve a spin flip of an electron on a single orbital: inelastic orbital excitations involve a transition to an orbital excited state. In the case of a magnetic atom with a half filled $d_z^2$ orbital **(a)** spin IETS and orbital IETS transitions are allowed for both tunnelling directions (bias polarities). In the case of a fully filled (or empty) $d_z^2$ orbital **(b)** spin excitations are blocked, and orbital excitations are only possible for tunnelling from substrate to tip, i.e. negative sample bias. **(c)** DFT calculated relaxed structure for Co monomer and **(d)** infinite double-spaced Co chain on $Cu_2N$. Calculated magnetization density is shown for **(e)** a Co monomer and **(f)** infinite chain.

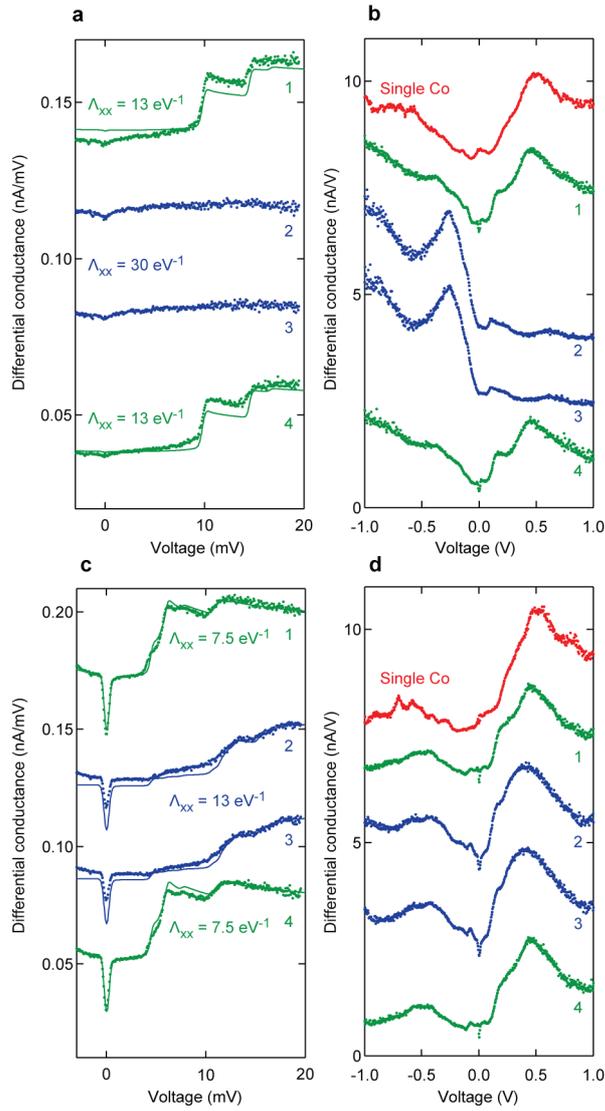

**Figure 4: Magnetic anisotropy of Co nanostructures (a)** IETS spectra of each atom of a Co tetramer built on Cu$_2$N. IETS steps on the inner atoms are suppressed, in the same way as in Fig. 1e. Solid lines are simulated spectra allowing the anisotropy parameter $\Lambda_{xx}$ to vary as indicated between the end and inner atoms. $\Lambda_{zz}$ = 6.0 eV$^{-1}$ and $\Lambda_{yy}$ = 0 in all cases, the antiferromagnetic spin coupling is 2.5 meV. For comparison, single Co may be fitted with $\Lambda_{xx}$ = 6.2 eV$^{-1}$. Simulated spectra are not shown for inner atoms since in these, the spin IETS transitions are suppressed. **(b)** Higher bias spectra on the same tetramer as (a). Note suppression of the positive bias spectral features in the inner atoms. **(c)** IETS spectra of each atom of another Co tetramer built on Cu$_2$N, in which spin excitations are not suppressed. This is thought to be due to local lattice strain. Simulated spectra are shown as for (a). The antiferromagnetic spin coupling is 2 meV. **(d)** Higher bias spectra on the same tetramer as **(c)**. The spectra are all observed to be similar to that of a single Co atom.

# Supplementary Material for:

# Controlled complete suppression of single-atom inelastic spin and orbital cotunnelling

B. Bryant[1], R. Toskovic[1], A. Ferrón[2,3], J. L. Lado[2], A. Spinelli[1], J. Fernández-Rossier[2], and A. F. Otte[1]

[1]Department of Quantum Nanoscience, Kavli Institute of Nanoscience, Delft University of Technology, Lorentzweg 1, 2628 CJ Delft, The Netherlands

[2]International Iberian Nanotechnology Laboratory (INL), Avenida Mestre José Veiga, 4715-310 Braga, Portugal

[3]Instituto de Modelado e Innovación Tecnológica (CONICET-UNNE), Avenida Libertad 5400, W3404AAS, Corrientes, Argentina

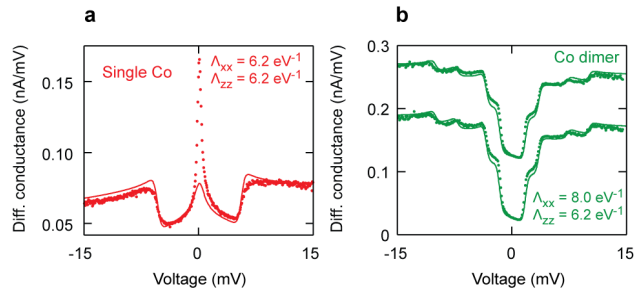

**Figure S1 Comparison of single Co and Co dimer spectra (a)** dI/dV spectrum of a single Co atom on $Cu_2N$ (dots). The solid line is a simulated spin IETS spectrum using $S = 3/2$, $\Lambda_{xx} = \Lambda_{zz} = 6.2$ $eV^{-1}$, $\Lambda_{yy} = 0$, equivalent to $D = 2.7$ meV, $E = 0$. Interactions with substrate electrons are taken into account up to 3$^{rd}$ order, using an antiferromagnetic Kondo-exchange coupling $J_K\rho_s = -0.15$ meV. However, this still underestimates the height of the Kondo peak. **(b)** dI/dV spectra of each atom of a Co dimer on $Cu_2N$ (dots). This is the same dimer as shown in Fig. 1. Solid lines are simulated spin IETS spectra including an antiferromagnetic spin coupling of $-2$ meV. $\Lambda_{xx} = 8.0$ $eV^{-1}$, $\Lambda_{zz} = 6.2$ $eV^{-1}$, $\Lambda_{yy} = 0$, $J_K\rho_s = -0.08$ meV. Importantly, only the anisotropy term along the dimer (*x*) axis is changed from the single atom case.

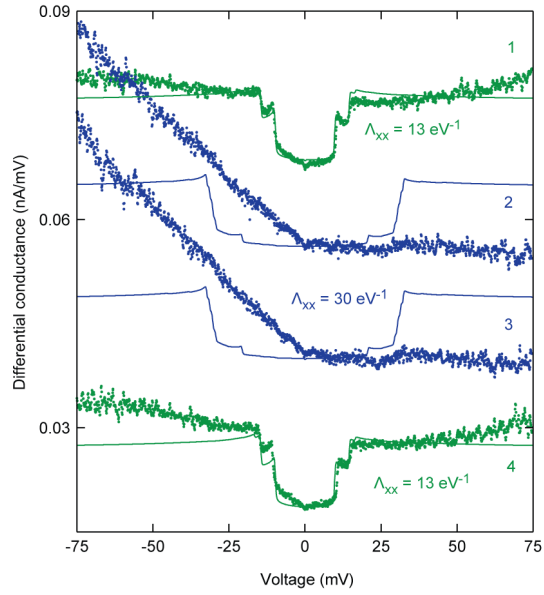

**Figure S2 Spectra of a Co tetramer up to 75 mV.** dI/dV spectra (dots) of a Co tetramer similar to Fig. 1c and Fig. 4a, at up to 75 mV. On the inner atoms, no spin excitation steps are seen. The solid lines are simulated spin excitation spectra. Although no spin excitations are observed on the inner atoms, it is necessary that the inner atom x-axis anisotropy $\Lambda_{xx} > 25$ eV$^{-1}$ to reproduce the end atom spectra. For the model shown, for all atoms $S = 3/2$, $J_K \rho_s = -0.05$ meV, $\Lambda_{zz} = 6.2$ eV$^{-1}$, $\Lambda_{yy} = 0$ and the antiferromagnetic spin coupling is $-2.5$ meV. For the end atoms $\Lambda_{xx} = 13$ eV$^{-1}$, for the inner atoms $\Lambda_{xx} = 30$ eV$^{-1}$. The model reproduces the end atom spectra accurately: for the inner atoms it predicts a step at 30 meV which is not observed, since spin excitations on these atoms are suppressed due to the strongly hybridised $d_z^2$ orbital.

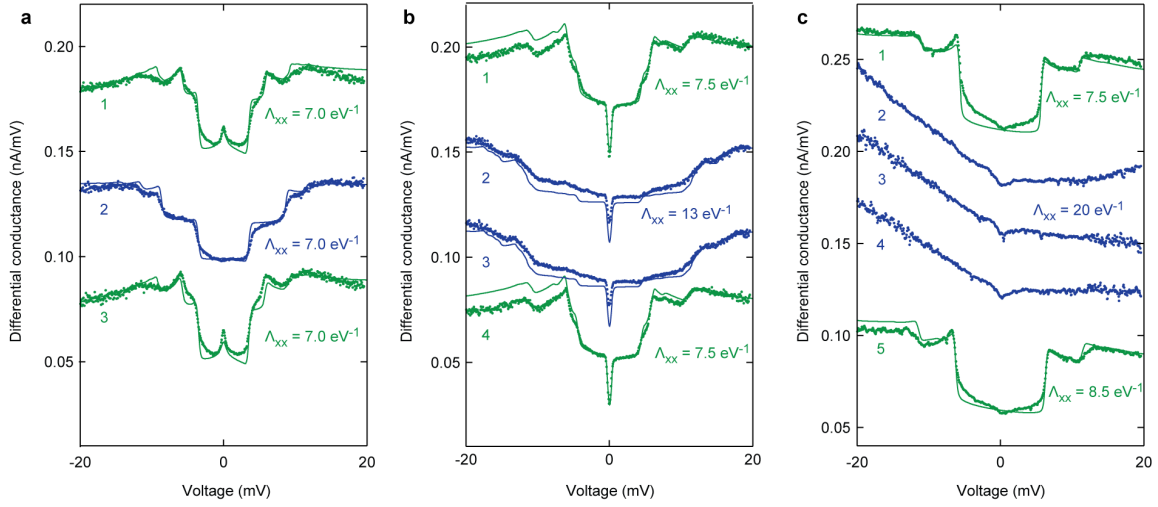

**Figure S3 Co nanostructures with suppressed and unsuppressed spin excitations.** In most cases, Co chains longer than two atoms built on $Cu_2N$ showed supressed spin excitations on the inner atoms (Fig. 1). In a minority of cases (4 out of 20 structures), Co chains up to four atoms long were observed to have unsuppressed spin excitations. This effect was specific to certain locations on the substrate, and is presumably due to variations in local strain induced by subsurface defects. **(a)** dI/dV spectra of a Co trimer and **(b)** a Co tetramer, as shown in Fig. 4c, constructed as an extension of the trimer, both of which show unsuppressed spin excitations. Simulated Spin IETS spectra are shown (solid lines), for all atoms $S = 3/2$, $\Lambda_{zz} = 6.2$ eV$^{-1}$, $\Lambda_{yy} = 0$. The trimer may be fitted with $\Lambda_{xx}= 7.0$ eV$^{-1}$ for all atoms: for the tetramer $\Lambda_{xx} = 7.5$ eV$^{-1}$ for the end atoms and 13 eV$^{-1}$ for the inner atoms. For both structures the antiferromagnetic spin coupling is −2 meV, $J_K\rho_s = -0.05$ meV for inner atoms and −0.15 meV for the end atoms. **(c)** dI/dV spectra of a Co pentamer constructed as an extension of (b). In this case no spin IETS steps are seen on the inner atoms: suppression of spin excitations was observed on *all* structures longer than four atoms. We may conclude that the strain induced by adding a Co atom to the structure overcomes the local lattice strain, pushing all inner atoms of the structure beyond the critical N-Co-N angle that leads to suppression of spin excitations. The end atom spectra may be simulated using a similar model to Fig. S2: for all atoms $S = 3/2$, $J_K\rho_s = -0.05$ meV, $\Lambda_{zz} = 6.2$ eV$^{-1}$, $\Lambda_{yy} = 0$, the antiferromagnetic spin coupling is −2.0 meV. For the end atoms $\Lambda_{xx} = 7.5$ and 8.5 eV$^{-1}$, for the inner atoms $\Lambda_{xx} = 20$ eV$^{-1}$. Note that the anisotropy values are smaller than in Fig. S2, illustrating the effect of the local strain.

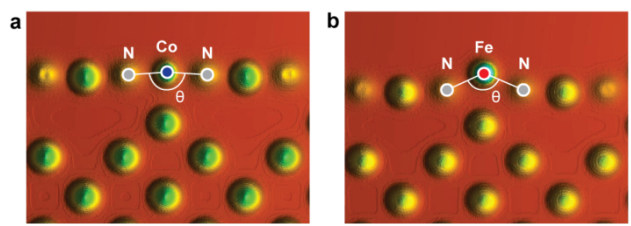

**Figure S4 Comparison of Co chains and Fe chains on Cu$_2$N (a)** section through DFT-computed relaxed structure of an infinite chain of Co atoms on Cu$_2$N, as shown in Fig. 3d. The N-Co-N angle θ is 175°, leading to strong hybridization of the $d_z^2$ orbital and suppression of spin excitations. **(b)** Relaxed structure of an infinite chain of Fe atoms on Cu$_2$N. The N-Co-N angle θ is much smaller at 133°, so in this case the $d_z^2$ orbital is not strongly hybridized and spin excitations are not suppressed.